# Effect of the Density of Reactive Sites in P(S-r-MMA) Film During Al₂O₃ Growth by Sequential Infiltration Synthesis

*Federica E. Caligiore, Daniele Nazzari, Elena Cianci\*, Katia Sparnacci, Michele Laus, Michele Perego\*, Gabriele Seguini.*

F. E. Caligiore, Dr. E. Cianci, G. Seguini, Dr. M. Perego, D. Nazzari

IMM-CNR,

Unit of Agrate Brianza,

Via C. Olivetti 2, 20864 Agrate Brianza (MB), Italy

E-mail: michele.perego@cnr.it,  elena.cianci@cnr.it

D. Nazzari now at TU Wien, Institute for Solid State Electronics, Gußhausstraße 25-25a, Wien, Austria.

Prof. K. Sparnacci, Prof. M. Laus

Dipartimento di Scienze e Innovazione Tecnologica (DISIT),

Università del Piemonte Orientale "A. Avogadro",

Viale T. Michel 11, 15121 Alessandria, Italy

Abstract.

Sequential infiltration synthesis (SIS) consists in a controlled sequence of metal organic precursors and co-reactant vapor exposure cycles of polymer films. Two aspects characterize a SIS process: precursor molecule diffusion within the polymer matrix and precursor molecule entrapment into polymer films *via* chemical reaction. In this paper, we investigated SIS process for the alumina synthesis using trimethylaluminum (TMA) and $H_2O$ in thin films of poly(styrene-*random*-methyl methacrylate) (P(S-*r*-MMA)) with variable MMA content. The amount of alumina grown in the P(S-*r*-MMA) films linearly depends on MMA content. A relatively low concentration of MMA in the copolymer matrix is enough to guarantee the volumetric growth of alumina in the polymer film. In pure polystyrene, metal oxide seeds grow in the sub-surface region of the film. *In-situ* dynamic spectroscopic ellipsometry (SE) analyses provide quantitative information about TMA diffusivity in pristine P(S-*r*-MMA) matrices as a function of MMA fraction, allowing further insight into the process kinetics as a function of the density of reactive sites in the polymer film. This work improves the understanding of infiltration synthesis mechanism and provides a practical approach to potentially expand the library of polymers that can be effectively infiltrated by introducing reactive sites in the polymer chain.

## 1. Introduction

Sequential infiltration synthesis (SIS) belongs to a family of vapor phase infiltration techniques and it is derived from atomic layer deposition (ALD) on polymers[1]. In SIS, the vapor precursor pulses are alternated by an exposure time orders of magnitude larger than that in ALD. This longer exposure allows the precursors to diffuse into the polymer substrates rather than simply decorating the surface[2–4]. SIS allows transforming polymers into organic-inorganic hybrid materials: penetration and reaction of gaseous metal precursors into polymer, during SIS, permit to grow inorganic materials into polymeric films[5] in order to tune some of their features as their optical properties or improve their chemical etch resistance[6–9].

The SIS process is characterized by two factors: diffusion and entrapment of precursor molecules into the polymer matrix. The most direct method to promote their entrapment is the chemical reaction of penetrant molecules with polymer functional groups in combination with a secondary precursor (co-reactant)[4]. Thus, the number of reactive sites within the polymer film plays an important role in the determination of the amount of inorganic material grown in the film during the process. Moreover, differently from ALD, the self-terminating reactions are not restricted to the surface sites. Precursor molecules must diffuse into the polymeric film to reach the reactive sites that are distributed in the volume of the polymeric matrix. Consequently, diffusion plays a fundamental part in the kinetics of the infiltration process and needs to be monitored in real time. The most used *in-situ* techniques for investigating the infiltration mechanism are *in-situ* quartz crystal microbalance (QCM) measurements and *in-situ* Fourier Transform Infrared (FTIR) spectroscopy[10–13]. *In-situ* dynamic spectroscopic ellipsometry (SE) was recently validated as a valuable tool for real time investigation of the infiltration process in PMMA and PS homopolymers[14]. *In-situ* SE is a non-invasive optical technique frequently used in combination with ALD processes[15–17] and for studies of polymer films under several conditions[18–20]. During SIS process, *in-situ* SE allows continuous acquisition of information about changes of thickness and refractive index ($n$) of polymer films without interrupting the process itself on a shorter time scale than *in-situ* FTIR spectroscopic analysis and without the need of *ad hoc* samples as for QCM measurements.

When SIS is performed into self-assembled block copolymers (BCP)[21], the selective binding of precursors to one domain only of BCPs offers the possibility to fabricate inorganic functional nano-architectures[22] or hard masks for lithography[23]. Removing polymers by $O_2$ plasma after infiltration yields inorganic nanostructures reproducing the selected polymer domain[2,3,6,7,24–30]. Various gas phase and liquid phase processes have been utilized to synthetize patterned nanomaterials using BCPs as scaffolds[31,32]. SIS overtakes these approaches in terms of reproducibility and possibility to up scaling. In fact, the high control of reaction during a SIS process permits to reproduce the morphology of the BCP template, finely tuning the dimension of the generated nanofeatures without changing the domain size of BCPs[3]. Infiltration of trimethylaluminum (TMA) in combination with $H_2O$ for the synthesis of $Al_2O_3$ into polystyrene-*block*-poly(methyl methacrylate) (PS-*b*-PMMA) films is the most studied SIS process[25,30]. TMA has a high reactivity at low temperature and interacts selectively with the carbonyl groups in the MMA units[3,25,33]. PS-*b*-PMMA thin films in combination with SIS have been used to fabricate inorganic $Al_2O_3$ nanostructures used as hard mask for pattern transfer[34–36].

In this respect, dots and stripes, two basic elements of lithographic patterns, can be obtained from BCP thin films[37–40] featuring cylinders and lamellae perpendicularly oriented with respect to the substrate[41]. Thus, the control of the domain orientation represents an essential key for the technological application of these materials in lithographic processes. In order to obtain perpendicular orientation, non-preferential interactions are necessary between polymer/substrate and polymer/air interfaces[22]. The most common strategy to achieve substrate neutrality involves the introduction of a brush layer of hydroxyl-terminated random copolymer (RCP)[42] with tailored composition[43]. P(S-*r*-MMA) RCPs are typically used to induce perpendicular orientation of phase separated PS-*b*-PMMA BCP thin films[43–48]. It is important to note that, during a SIS process in perpendicularly oriented PS-*b*-PMMA BCP thin films, the underlying P(S-*r*-MMA) brush layer is infiltrated too, creating a continuous alumina layer on the $SiO_2$ layer of silicon substrate. Consequently, in order to optimize infiltrated BCP nanoarchitectures for lithographic use, an in-depth study of the infiltration mechanism in RCPs films is mandatory. Moreover, from a general point of view, to the best of our knowledge, no data are available in the literature about the kinetics of the SIS process as a function of the density of reactive sites in the polymer matrix. More detailed information about the intrinsic correlation between the growth of $Al_2O_3$ and the density of reactive sites in the polymer could help to unveil the fundamental mechanism governing the SIS process, providing the possibility to increase our growth capability and expand its technological applicability.

In this work, we investigated the infiltration process of TMA and $H_2O$ in P(S-*r*-MMA) films with a variable MMA unit content ranging from 12% to 77%, using *ex-situ* SE and *in-situ* dynamic SE. By *ex-situ* SE, we analyzed alumina layer growth in P(S-*r*-MMA) RCP films upon removing the polymer matrix. We explored how the variation of MMA unit content within the polymer film influences the growth of alumina. *In-situ* dynamic SE analyses provided quantitative information about TMA diffusivity in pristine P(S-*r*-MMA) matrices and further insights into the process kinetics as a function of the density of reactive sites in the polymeric film.

## 2. Results

### 2.1. Ex-situ analysis: Al$_2$O$_3$ grown in 10-SIS-cycles

**Figure 1(a)** shows the alumina film thickness obtained at the end of 10 cycles of infiltration on 55 nm thick RCP, PMMA, and PS films as a function of xMMA, after the selective removal of the organic component by O$_2$ plasma treatment. Al$_2$O$_3$ film thickness, measured by *ex-situ* SE, shows a linear relationship with the MMA unit fraction value. In **Figure 1(b)**, the initial thickness of polymer films are reported, showing a limited variation from sample to sample. Consequently, the observed linear evolution of the Al$_2$O$_3$ film thickness exclusively depends on the film composition and is directly related to the concentration of reactive sites in the polymer matrix. As a matter of fact, by increasing MMA concentration in P(S-*r*-MMA), the number of reactive C=O groups in the volume of the polymer film progressively increases. In particular, the experimental data suggest that the diffusion of TMA is fast enough to completely fill the volume of the 55 nm thick P(S-*r*-MMA) and PMMA films and that the amount of Al$_2$O$_3$ grown into the polymeric film during the SIS process is essentially limited by the number of reactive sites in the system. Actually, a not negligible amount of alumina is present on the substrate upon removal of the PS matrix, indicating a limited but not negligible incorporation of Al$_2$O$_3$ in the PS matrix, consistently with data available in the literature[1,49–51].

Refractive index and porosity of SIS Al$_2$O$_3$ are reported as a function of MMA fraction in **Figure 2**. Refractive indices are extracted at a wavelength of 632.8 nm and are roughly constant irrespective of xMMA. Despite some dispersion of the data, it is worth to note that the average value (n=1.50) of the refractive index is lower than the one of alumina grown by conventional ALD at 90°C, *i.e.* n=1.54[52,53]. The lower value of the refractive index corresponds to a lower density of $Al_2O_3$ films obtained by SIS compared to ALD $Al_2O_3$ and indicates that these films are slightly porous. The average porosity of SIS $Al_2O_3$ films is around 10%. Refractive index and porosity of alumina layers are independent of the MMA unit fraction. Additionally, the refractive index of the metal oxide film obtained by infiltration of the PS film is much lower than those obtained by infiltration of P(S-*r*-MMA) films. In this case, the resulting alumina film exhibits refractive index n=1.36, corresponding to a porosity of 31%. **Figure 3** reports the SEM images of $Al_2O_3$ layers obtained from infiltration of PMMA, P(S-*r*-MMA) (xMMA=0.40), and PS after removal of the polymer matrix by $O_2$ plasma. Actually, in case of PMMA and P(S-*r*-MMA) the morphology of alumina films are comparable, showing a compact and uniform $Al_2O_3$ surface. Conversely, alumina grown in PS film exhibits a completely different morphology, in agreement to the very high porosity of the $Al_2O_3$ film determined by analysis of the ellipsometric data and with our previous results[14]. Interestingly, the presence of MMA units in the polymer chain significantly modifies the morphology of the $Al_2O_3$ film obtained by SIS, suggesting that the presence of MMA in the volume of the polymeric film induces a different initial growth mechanism of $Al_2O_3$ into the polymeric film, even for very small MMA concentration. In order to get more information on this specific issue, **Figure 4** reports the evolution of $Al_2O_3$ film thickness upon 10 SIS cycles and $O_2$ plasma treatment as a function of the initial thickness of the polymeric film. In particular, in **Figure 4(a)** $Al_2O_3$ obtained from SIS in PS films is compared to that obtained from PMMA films. Interestingly $Al_2O_3$ thickness increases linearly as a function of PMMA initial thickness, suggesting that, in the range of thickness we considered, alumina grows in the whole PMMA volume. Conversely, in PS films, the thickness of $Al_2O_3$ increases linearly when the PS film thickness is below or equal to 50 nm and then reaches a constant value. Thus, in PS samples, we can argue that the alumina growth occurs only within the sub-surface of the polymer film with an infiltration depth of approximately 50 nm. This picture is perfectly consistent with previous experimental results reported in the literature[1,49–51]. **Figure 4(b)** reports the evolution of $Al_2O_3$ film thickness as a function of the initial thickness of the polymeric film for the P(S-*r*-MMA) RCPs with different MMA content. Similarly to PMMA case, alumina film thickness increases linearly as a function of the polymer initial thickness.

In **Figure 4(c)**, the ratio between alumina thickness grown in 10 SIS cycles and polymer film initial thickness ($h_0$) is plotted as a function of xMMA. It increases linearly with the MMA percentage. This result suggests that the amount of TMA that reacts in the volume of the polymer film is proportional to the amount of MMA in the film itself. The slope of the linear fit obtained for low PS initial thickness is clearly out of trend.

## 2.2. In-situ analysis: the first SIS cycle

By means of *in-situ* SE analysis it is possible to measure thickness variation of the polymeric film during the SIS process in real time and to obtain information about TMA diffusion kinetics during the infiltration process[14]. Analysis of *ex-situ* data indicates that the amount of $Al_2O_3$ that is grown per unit volume in a P(S-*r*-MMA) matrix is independent of the thickness of the polymer film for each specific MMA fraction. Consequently, in following paragraphs, *in-situ* measurements will focus on the 55 nm and 100 nm thick samples to increase the accuracy of the measurements without any loss of generality, consistently with data reported in our previous publication[14].

**Figure 5** shows the thickness variation of polymer films in the first SIS cycle as a function of time for 55 nm and 100 nm thick P(S-*r*-MMA) RCP films with and MMA fraction ranging from 12% to 77%. During the first cycle, the polymer thickness increases during TMA injection and decreases during the following TMA purge and $H_2O$ pulse, similarly to what already reported for PMMA films[14]. The amplitude of film swelling depends on the initial thickness and on the percentage of MMA inside the polymeric matrix. 100 nm thick films go through a larger swelling than 55 nm thick films. Moreover, swelling decreases progressively by reducing xMMA. This result perfectly agrees with a model that links polymer thickness variation with the capability of the precursor to react with specific functional sites corresponding to C=O groups of the MMA units. In particular, the availability of more functional sites means more precursor molecules forming a C=O···Al(CH$_3$)$_3$ adduct. This larger amount of reacted precursor determines the larger thickness variation. In other words, the first cycle is characterized by a composition dependent thickness evolution evidencing the different interaction of TMA in each polymer and copolymer depending on MMA content. From a general point of view, we can consider TMA acting as a solvent for polymers with MMA reactive sites. Consequently, we expect the maximum swelling ratio to be directly proportional to the concentration of solvent, to the polymer-solvent interaction strength, and to the elastic properties of the polymer matrix.

To investigate the kinetics of the TMA diffusion process we define the polymeric film swelling $\varepsilon(t)$ as:

$$\varepsilon(t) = \frac{h(t) - h_0}{h_0} \tag{1}$$

where $h_0$ and $h(t)$ are the thickness of polymer film at time $t=0$ and $t$ respectively[54,55]. **Figure 6** reports the normalized swelling $\varepsilon(t)/\varepsilon_{max}$ as a function of $\sqrt{t}/h_0$ for 55 and 100 nm thick films with different composition. $\varepsilon_{max}$ corresponds to the maximum swelling of the polymeric film during the precursor exposure. All the curves follow a linear trend for small $t$ values, proving that TMA infiltration is governed by a Fickian diffusion process[56] during the initial stage of TMA exposure for all film composition, independently of film thickness. Starting from these data, it is possible to derive TMA diffusion coefficients (D) as the slope of the linear fitting of the swelling curves, according to the following equation[57]:

$$\varepsilon(t) \sim 4\varepsilon_{max}\sqrt{\frac{D}{\pi}}\sqrt{\frac{t}{h_0^2}} \qquad\qquad (2)$$

where D is an "effective" diffusion coefficient of TMA molecules in the specific matrix under investigation. Actually, this effective diffusion coefficient takes into account both diffusion and reaction kinetics of precursor molecules in the polymer. For fast reaction rates, the diffusion behavior can be described as Fickian, but with a reduced diffusion constant $D = D_0/(R+1)$ being R the ratio of immobilized precursor to freely diffusing precursor and $D_0$ is the diffusion coefficient in the case of purely diffusional kinetics[4].

**Figure 7** shows the effective diffusion coefficients calculated as a function of xMMA. D progressively increases with xMMA for xMMA<0.56 reaching a maximum value at xMMA=0.56. For further increase in xMMA, we observe a progressive decrease of D. This decrease of the effective diffusivity value for large xMMA values can be easily rationalized in the framework of the effective diffusion coefficient model[4]. In more details, the increase in the concentration of reactive sites in the polymeric matrix implies larger R values and a reduction of D values, with a significant slowing down of the global kinetics of the SIS process. However, based on the data obtained at small xMMA, we can conclude that other parameters have to be taken into account in order to explain the evolution of D as a function of xMMA. We can argue that the diffusion of TMA in P(S-*r*-MMA) thin films is significantly influenced by the variation of free volume in the polymeric matrix. In fact, PS films are much less porous than PMMA films[49]. When introducing MMA units in PS matrix, the effective diffusion coefficient is determined by a delicate balance between the increase in diffusivity, which is associated to the larger free volume, and the decrease of diffusivity, that is determined by the increased reactivity (larger R) of the polymeric matrix. At small xMMA values, the effective diffusivity progressively increases with xMMA due to the larger free volume in the polymeric matrix. For further increase in MMA content, the reaction with C=O groups dominates and slows down the TMA diffusion in the polymeric matrix, resulting in a progressive reduction of D as a function of xMMA, in agreement with the model described by Leng and Losego[4].

By means of these effective diffusion coefficients, it is possible to model the evolution of the TMA concentration profile in the polymeric matrix as a function of time. In particular, using concentration profile equation for a finite plane slab, with constant concentration at the surface ($C_s$)[57], it is possible to calculate the TMA concentration profile upon 60 s of TMA exposure and to determine if TMA can diffuse through the entire film thickness h in the examined polymeric films, reaching a saturation level. The equation and related concentration profiles are reported in Supporting Information (Figures S1 and S2). **Figure 8** reports the $C_h/C_s$ as a function of xMMA for the 100 and 55 nm thick films. $C_h$ represents the TMA concentration at the polymer/substrate interface after diffusion through the whole film thickness (h). Saturation corresponds to $C_h$ equal to $C_s$. In the case of 55 nm thick films, $C_h/C_s$ values are approximately equal to 1 for all the polymers irrespective of xMMA value. Conversely, in the 100 nm thick samples, the $C_h/C_s$ values are well below the saturation level for all xMMA values.

The case of PS films deserves specific considerations, since no reactive groups are present and low free volume is available to TMA diffusion, resulting in low diffusion and reaction into the polymer film. Experimental data indicate a very small thickness variation of PS films during the first TMA exposure, as shown in the inset of **Figure 9**. From these data, we estimated a diffusion coefficient for TMA in PS films from the slope of the linear fit of the normalized swelling $\varepsilon(t)/\varepsilon_{max}$ $vs$ $\sqrt{t}/h_0$ (**Figure 9**). The corresponding D value was extrapolated to be $4.09 \times 10^{-13}$ cm²/s. The limited swelling that was observed in the PS films reduces the accuracy of the procedure. However, this value is smaller than those reported in **Figure 7** and it follows the outlined trend of a progressive reduction of TMA diffusivity when reducing xMMA in the polymeric matrix. Using this diffusion coefficient, we calculated the expected TMA concentration finding that PS films of 100 and 55 nm were filled up to 30% and 89% respectively. It is interesting to notice that, according to this simple model, the 55 nm thick PS film is too thick to be saturated by TMA during a 60 s long TMA exposure. This result perfectly matches the data reported in **Figure 4**, further confirming that TMA diffusion and, consequently, alumina growth in PS films are limited to the sub-surface region of the polymeric layer.

During the purging step, the reactor is vacuum pumped and then crossed by 100 sccm $N_2$ flow for 60 s. A significant thickness decrease is observed during the purging step of the first SIS cycle. **Figure 10** reports the thickness variation as a function of time for the 55 nm thick RCP films during the purging phase. The thickness evolution as a function of time follows an exponential trend $e^{-t/\tau}$. $\tau$ is assumed to be the time constant of the de-swelling process associated to the out-diffusion of non-reacted TMA[14]. Actually, competing processes with different characteristic times occur simultaneously during the de-swelling process. In addition to the loss of TMA by out-diffusion, the polymer relaxation can further contribute to the observed decrease of the film thickness[58]. **Figure11** reports the $\tau$ values as a function of MMA fraction into the film calculated for the 55 nm thick polymeric films. It is worth to note that PS de-swelling data does not properly follow an exponential decay, preventing the possibility to determine a proper $\tau$ value for this polymeric film. For all the other samples, $\tau$ values are independent of RCP composition. As reported in our previous work[14], $\tau$ values obtained by *in-situ* SE are found to be much shorter than those evaluated from FTIR analysis[11,12]. Actually, $\tau$ values obtained monitoring C=O absorbance peak intensity variation are of the order of tens of minutes and they were attributed to the desorption of physisorbed TMA from the reversible C=O···Al(CH$_3$)$_3$ adducts that are formed during TMA exposure phase[11,12]. By *in-situ* dynamic SE the de-swelling process of the polymer is observed during the elimination of the TMA vapor from the chamber and purging in $N_2$ flow. Therefore, the observed thickness evolution was mainly associated to the out-diffusion of non-reacted TMA molecules from PMMA films. We argue that the results on $\tau$ values can be related to how much the samples have been filled during the exposure phase.

In **Figure12** we report the values of the thickness variation of 55 nm thick samples measured at different stages during the first cycle of the SIS process, namely at the swelling peak ($h_1$), at the end of TMA purge ($h_2$) and after $H_2O$ injection ($h_3$). Data are reported as a function of xMMA. Interestingly, $h_i$ values increase linearly with xMMA. Considering that in 55 nm thick samples, irrespective of the diffusion coefficient, a 60 s long exposure time is sufficient to reach the saturation, the collected data clearly indicate that the amount of TMA trapped in the film during the first infiltration cycle is proportional to the number of reactive sites in the polymer films. Moreover, the linear evolution of $h_2$ and $h_3$ values as a function of xMMA suggests that the amount of $Al_2O_3$ in the polymeric film at the end of the cycle is directly proportional to the number of MMA reactive sites in the polymer matrix and not influenced by the variations of TMA diffusivity.

**3. Discussion**

The combination of *ex-situ* and *in-situ* analysis provides a comprehensive picture of SIS process in P(S-*r*-MMA) copolymer with MMA content ranging from 0 to 100%. As clearly evidenced in Fig. 1-3, alumina layers grown by SIS in 55 nm thick P(S-*r*-MMA) RCPs present a linear increase of their thickness as a function of the MMA fraction in the polymer film. Moreover, they exhibit the same morphology suggesting the formation of a homogenous $Al_2O_3$ film, as in the case of the $Al_2O_3$ films grown in pure PMMA films. Conversely, if there are no functional MMA groups in the polymeric film, as in the case of PS films, the infiltrated $Al_2O_3$ layer is very thin. Additionally, SEM inspection indicates a rough and porous morphology, suggesting a different growth mechanism in PS polymeric matrix, where the reactive sites are not C=O groups but defects or hydroxyl groups reached by the precursor in its short diffusion path. These data indicate that, in the presence of MMA units in the polymer matrix, the amount of $Al_2O_3$ grown during the SIS process is directly proportional to the concentration of MMA units in the RCP chains. This picture is perfectly consistent with previous experimental results showing that C=O functional groups in the MMA units act as seeds for the growth of $Al_2O_3$ during the SIS process[3,4,11–14,25].

This interpretation is further confirmed by the investigation of the amount of $Al_2O_3$ grown in the different polymeric matrices as function of their initial thickness. Interestingly, as shown in Fig. 4, infiltrated alumina thickness increases linearly as a function of polymer initial thickness for RCP and PMMA films, supporting the idea that the infiltration process occurs in the whole volume of the film at least for the range of thicknesses under investigation. Once again, it is worth noting that the behavior of PS films differs from PMMA and RCPs films, highlighting that the infiltration process occurs in a limited sub-surface region of the PS film.

*In-situ* analysis corroborates this picture. The effective TMA diffusion coefficient in RCP and PMMA films clearly indicates that, during the 60 s long exposure, TMA molecules diffuse through the entire volume of the 55 nm thick films. Therefore, this experimental finding clarifies that the linear dependence of $Al_2O_3$ film thickness on the xMMA in the 55 nm thick films is not associated to a partial filling of the polymeric film due to a variation of TMA diffusion coefficient with xMMA. The combination of *ex-situ* and *in-situ* analyses undoubtedly demonstrates that the final amount of $Al_2O_3$ in the polymeric film is essentially determined by the number of functional sites in the polymeric matrix.

In this respect, it is worth noting that the TMA diffusion coefficient we measured for the RCP and PMMA are fairly consistent with previous data reported in the literature for some molecules diffusing in a PMMA matrix[59]. A progressive scaling of the diffusivity is observed as a function of the size of the diffusing molecules[59]. Recent results about the infiltration in PMMA of $TiCl_4$ molecules, which has similar size of TMA, reports a diffusivity of $0.7 \times 10^{-12}$ $cm^2/s$ consistently with this picture[60]. Conversely, *ex-situ* analysis by Leng and Losego provided significantly different diffusion coefficients for TMA molecules in PMMA[61]. In particular, at 90°C they reported TMA diffusion coefficients in PMMA that are two orders of magnitude smaller than the one we obtained. Although we cannot exclude effects associated to different experimental set-up and processing conditions, we believe that they significantly underestimated the effective TMA diffusion coefficient and that the *ex-situ* analysis they performed does not provide a reliable protocol for an accurate measurement of the diffusion coefficient of precursors in a polymeric matrix during the SIS process. It is important to note that a $10^{-14}$ cm$^2$/s diffusion coefficient could not account for the linear dependence of infiltrated $Al_2O_3$ film thickness on the initial polymeric film thickness we observed, as TMA would not be able to fill the entire volume of the polymeric films in the time scale we considered in our experiments. The data of TMA diffusivity in PS films and the level-off of the $Al_2O_3$ thickness for PS films (Fig. 4) thicker than 50 nm are perfectly consistent with this picture and further support the idea that the TMA diffusion coefficients obtained by an *in-situ* approach provide a better description of the system. It is worth to note that these data could be extremely important for the modeling of SIS processes with standard simulation tools, providing the capability to effectively predict the characteristic features of the synthetized nanostructured materials when performing SIS in self-assembled BCP thin films.

Looking at the reported data from a more general perspective, we observe that the introduction of a very small fraction of MMA units is enough to promote the growth of alumina in the volume of polymer films. In fact, it is evident that even the presence of 12% of MMA causes a big change in the thickness and morphology of the resulting alumina films after 10 SIS cycles. We can argue that adding a very small fraction of MMA in a polymeric chain would make possible to functionalize it in order to obtain a volumetric growth of alumina by SIS in materials that usually do not provide evidence of efficient $Al_2O_3$ infiltration. Further study are currently on going to understand the minimum concentration of MMA that is necessary to achieve this effect.

Furthermore, from a technological point of view, this could pave the way to the application of SIS process to polymers other than PMMA. Interestingly, Kamcev *et al.* showed that it is possible to chemical modify self-assembled BCP thin films by ultraviolet light enhancing the block-selective affinity of organometallic precursors[26]. According to the authors, their results not only expand the functionality of PS-*b*-PMMA template but also suggest a general strategy for developing and identifying new BCP materials suitable for block-selective incorporation of a broader range of organometallic systems than currently possible. Their approach is based on a selective modification of the polymer structure by UV irradiation and consequently its application is limited by the effective capability to induce the required chemical modification only in the desired block, avoiding any degradation of the organic phase. Conversely, our data suggest the possibility to properly engineer the BCP chemical structure by introducing a small fraction of MMA reactive sites in one of the two blocks to promote selective growth of the inorganic phase in this specific block. The introduction of functional groups in one of the two blocks during the synthesis of the BCP offer the possibility to independently tune the properties of the BCP in terms of incompatibility between blocks and chemical selectivity for the SIS process. In particular, we could envision the possibility to grow very small inorganic nanostructures in high incompatibility BCPs modifying one of the blocks and promoting the selective growth of the inorganic phase, but preserving the high incompatibility between the two blocks of BCPs and guaranteeing the phase separation of BCP on length scale smaller than the ones achievable with a conventional PS-*b*-PMMA BCP.

## 4. Conclusion

P(S-*r*-MMA) RCPs thin films with different MMA fraction and thicknesses ranging from 8 to 100 nm were infiltrated using SIS process based on TMA and $H_2O$ precursors. $Al_2O_3$ layers obtained after removal of the polymer matrix by $O_2$ plasma were analyzed by *ex-situ* SE. The morphology and the density of the metal oxide layer is the same irrespective of the MMA content in the polymer matrix, while the final thickness of infiltrated alumina depends on the MMA fraction. *In-situ* SE investigations allowed a real-time monitoring of the SIS process, highlighting that TMA diffusion constant in the polymer film slightly depends on xMMA.

Collected data clearly indicate that the amount of $Al_2O_3$ grown by the SIS process is essentially determined by the availability of reactive sites within the polymeric film. These experimental results provide fundamental information about the intimate chemical-physical mechanisms governing the SIS process and suggest the possibility to explore chemical modifications of polymer chains by introduction of a small fraction of reactive sites, in order to enhance the reactivity of metallic precursors to the polymer matrix. From this point of view, this work advances our understanding of the SIS process and expand capability to control the growth of inorganic materials into an organic phase, providing interesting perspectives for the technological implementation of this process in BCP materials other than PS-*b*-PMMA.

## 5. Experimental Section

*P(S-r-MMA) RCP synthesis.* PS and PMMA homopolymers were purchased from Polymer Source Inc. and used without further purification. α-hydroxyl ω-Br functional RCPs were obtained by ARGET ATRP of styrene and methyl methacrylate initiated by HEBIB and catalyzed by $CuBr_2$/$Me_6$TREN complex in the presence of $Sn(EH)_2$ as the reducing agent[62]. By varying the molar ratio between styrene and methyl methacrylate, and keeping all other reaction parameters constant, copolymers with different composition were obtained. In addition, the reactions were stopped after a yield of approximately 20% were reached, to avoid or at least reduce composition dispersion effects due to differences in reactivity ratios. In this way, $M_n$ comprised between 10600 and 14800 g·mol$^{-1}$ with relatively narrow polydispersity index (Đ) are obtained.

Table 1 summarizes the fundamental characteristics of the different homopolymers (PS and PMMA) and P(S-r-MMA) RCPs used in this work. Our analysis was done as a function of xMMA, which is defined as the ratio between the number of MMA units ($N_{MMA}$) and the number of total monomer units constituting each P(S-*r*-MMA).

*Sample preparation for SIS.* Thin polymeric films ranging from 8 to 100 nm, were prepared by spin coating, properly adjusting the concentration of polymer-toluene solutions. (100) Si substrates were cleaned in Piranha solution ($H_2SO_4/H_2O_2$, 3/1 vol. ratio) at 80°C for 40 min in order to enhance the surface density of hydroxyl groups, rinsed with 2-propanol in ultrasonic bath and $N_2$ dried. Then, polymeric solutions were deposited and, subsequently, the films were thermally annealed at 250°C for 900 s in $N_2$ atmosphere by means of a Rapid Thermal Process (RTP) tool in order to remove liquid toluene in spinned film[47,63,64].

*SIS process.* Samples were loaded in a commercial cross flow ALD reactor (Savannah 200, Ultratech Cambridge NanoTech.) and thermalized at 90°C for 30 min under 100 sccm $N_2$ flow at 0.6 Torr before starting the infiltration. The samples underwent a 10-SIS-cycle process. TMA and $H_2O$ were the metal precursor and the oxidant respectively. Each SIS cycle consisted in successive pulses of TMA and $H_2O$, each one followed by an exposure step during which the system was isolated from the pumping line and samples were exposed to the precursor or oxidant vapor. A purging phase followed each exposure phase, during which the samples were exposed to 100 sccm of $N_2$ flow. The SIS cycle was 0.025 s TMA pulse/60 s exposure/60 s purge, followed by 0.015 s $H_2O$ pulse/60 s exposure/180 s purge. Upon the SIS process, samples were exposed to $O_2$ plasma to remove the polymer matrix, obtaining the alumina thin films on the substrate.

*Characterization.* *Ex-situ* and *in-situ* SE were performed using a rotating compensator ellipsometer equipped with Xe lamp (M-2000F, J. A. Woolam Co. Inc). For the *in-situ* investigation, the reactor lid was modified with two quartz windows to permit the incoming light to be sent onto the sample and the reflected beam to be detected, at 70° fixed angle with respect to the substrate plane normal. Ellipsometric data were collected over the wavelength range from 250 to 1000 nm throughout the entire SIS process. The acquisition time was of 2.5 s. However, as polymer modifications when TMA or $H_2O$ are injected in the ALD chamber are very fast, the acquisition time was reduced to 1.6 s during the single first infiltration cycle in pristine polymeric layer. EASE software package 2.3 version (J. A. Woollam Co. Inc.)[65] was used to analyze data and evaluate film thickness and refractive index. Ellipsometric data were fitted using a film stack model composed of a Cauchy layer model on 2 nm thick $SiO_2$ on the silicon substrate[66]. After $O_2$ plasma, the spectra of final alumina layers were collected by *ex-situ* SE with the same ellipsometer at fixed 75° incidence angle and were analyzed with the same stack model. In order to take into account the porosity of the alumina film, the *ex-situ* SE data were modeled also using the Bruggeman effective medium approximation (EMA).

Finally, the morphology of alumina layers was pictured by field emission scanning electron microscopy (FE-SEM, SUPRA 40, Zeiss) using an in-lens detector and an acceleration voltage of 15 kV.

**Supporting Information**

Concentration profile equation in case of an infinite plane slab and constant concentration at the surface and related concentration profiles. Fig. S1 and Fig. S2 report concentration profiles of TMA diffusivity in P(S-*r*-MMA) RCPs during 60 s of exposure for 100 and 50 nm thick samples respectively.

**Acknowledgements**
The authors want to acknowledge Mario Alia (CNR) for technical assistance, V. Gianotti, and D. Antonioli (Università Del Piemonte Orientale, Italy) for RCP synthesis. This research has been partially supported by the project "IONS4SET" funded from the European Union's Horizon 2020 research and innovation program under Grant No. 688072.
The manuscript was written through contributions of all authors. All authors have given approval to the final version of the manuscript.

Received: ((will be filled in by the editorial staff))
Revised: ((will be filled in by the editorial staff))
Published online: ((will be filled in by the editorial staff))

3920.

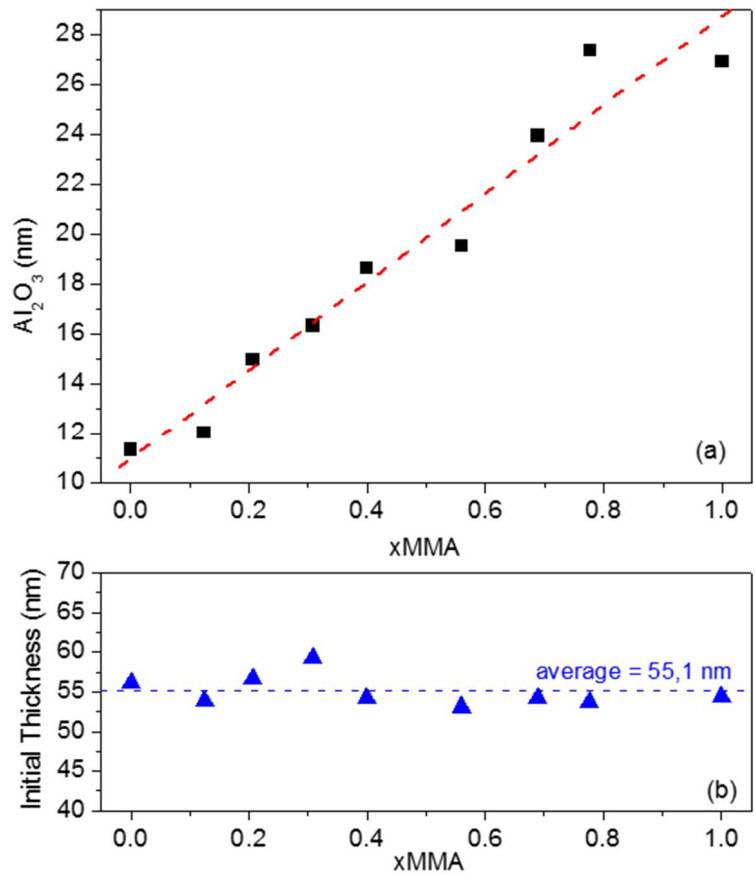

**Figure 1. (a)** Thickness of $Al_2O_3$ layer grown by 10 cycles of infiltration of ~55 nm thick P(S-*r*-MMA), PMMA, and PS samples as a function of xMMA. Alumina thickness grows linearly with MMA unit fraction. **(b)** Initial thickness of polymer films.

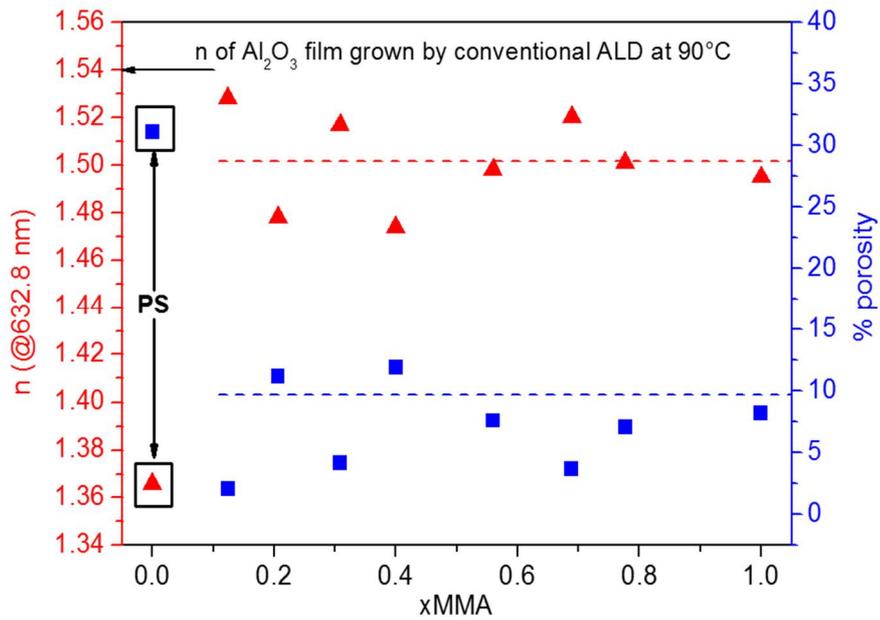

**Figure 2.** Refractive index n (red triangle dots) and porosity (blue squared dots) of alumina layer grown from SIS in 55 nm thick samples as function of xMMA. Refractive index and porosity values do not depend on MMA unit fraction; only in the case of PS, alumina film has a completely different value for n and porosity.

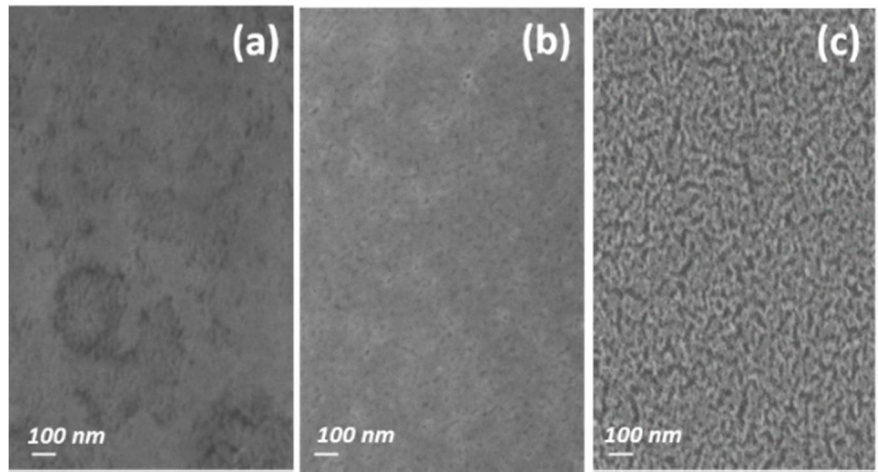

**Figure 3.** FE-SEM images of alumina films resulting from 10-SIS-cycle TMA/$H_2O$ SIS process in 55 nm thick PMMA **(a)**, P(S-*r*-MMA) (xMMA=0.40) **(b)**, and PS **(c)** after removal of polymer matrix in $O_2$ plasma.

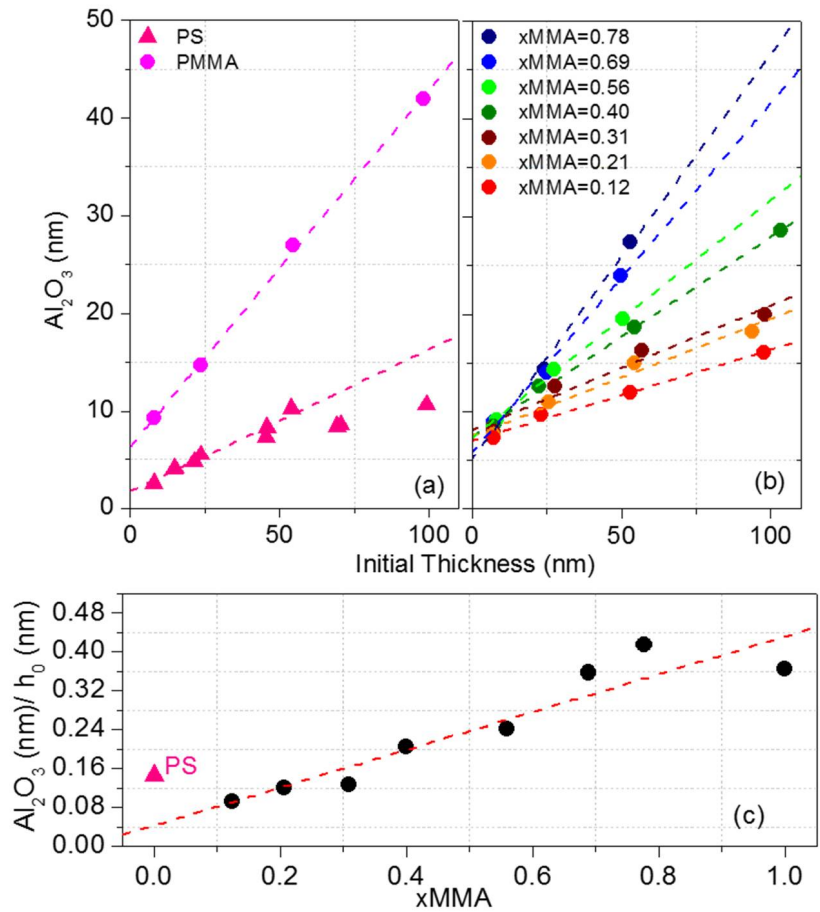

**Figure 4.** Thickness of the Al₂O₃ layers resulting from infiltration of PS, PMMA **(a)** and P(S-r-MMA) **(b)** samples as a function of polymer initial thickness. **(c)** Al₂O₃ to polymer thickness ratio obtained as slope of linear fits in (a) and (b) as a function of xMMA within the film. Pink triangle dot is the slope of linear fit for PS.

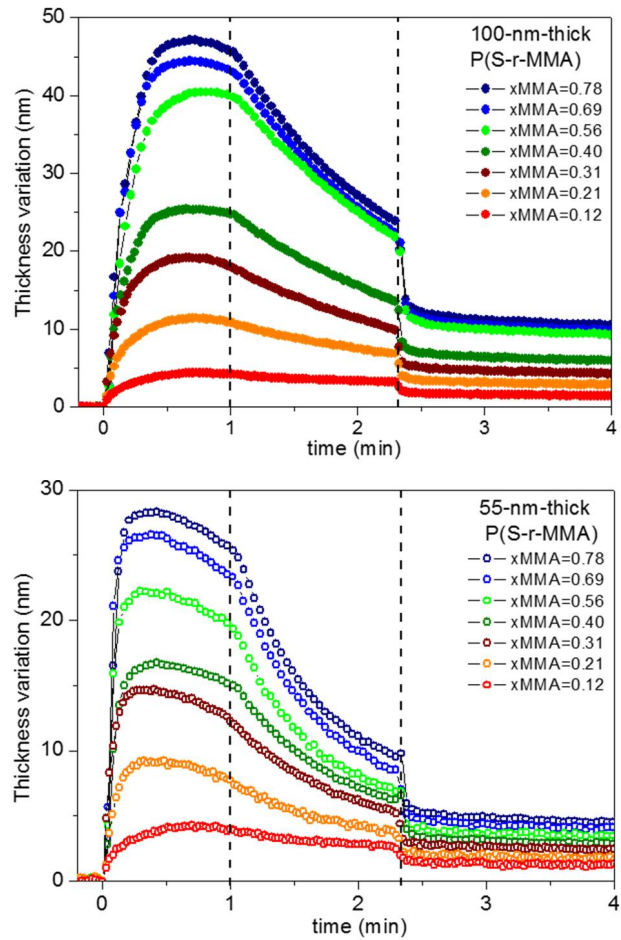

**Figure 5.** *In-situ* SE measured thickness variation of 100 and 55 nm thick films of P(S-*r*-MMA) RCPs during the first SIS cycle. The dashed line on the left defines the TMA exposure phase while the dashed line on the right marks the injection of $H_2O$ in the chamber.

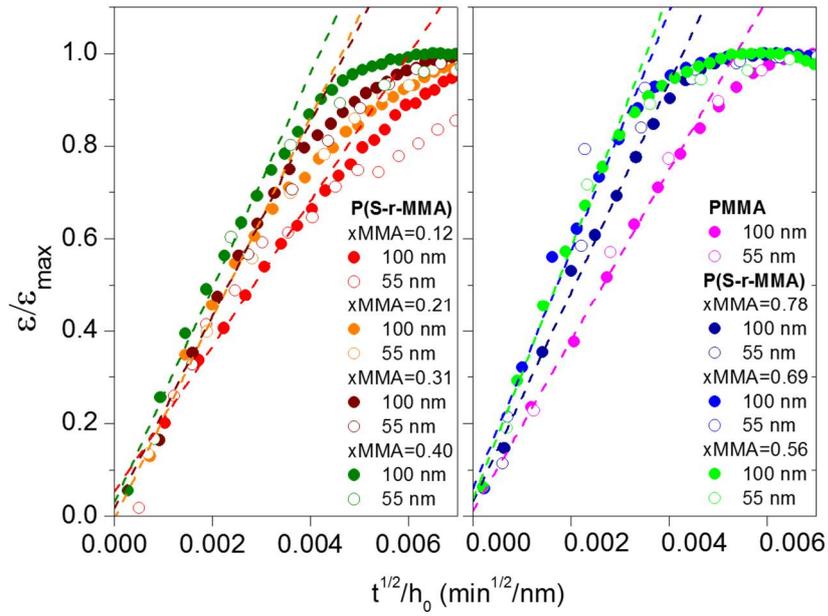

**Figure 6.** $\varepsilon(t)/\varepsilon_{max}$ during the 1st TMA exposure phase as a function of $\sqrt{t}/h_0$ for 55 and 100 nm thick P(S-*r*-MMA) samples with all investigated compositions. Data for PMMA films are also reported.

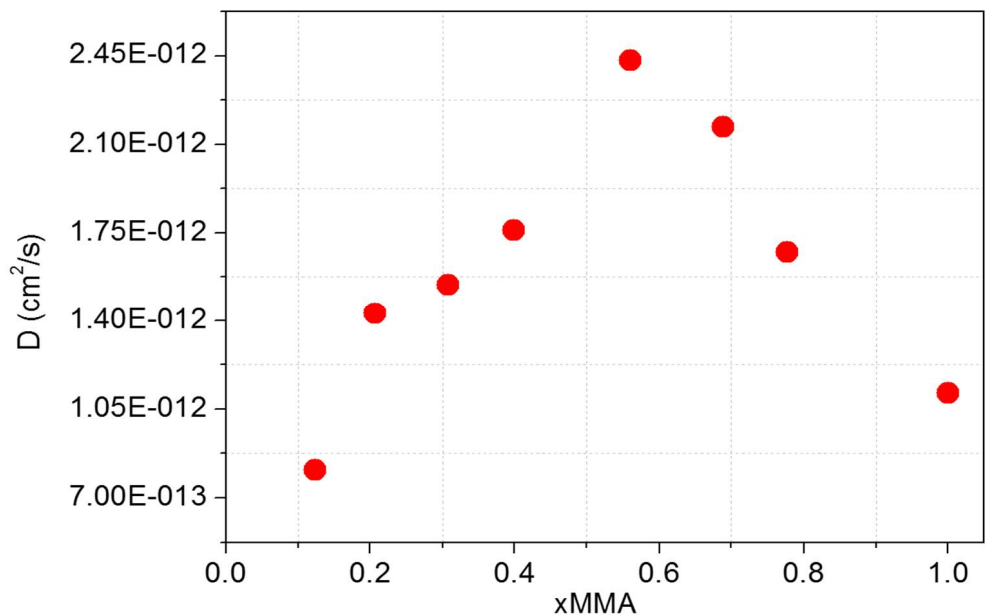

**Figure 7.** In-diffusion coefficients of TMA at 90°C in 100 and 55 nm thick P(S-r-MMA) RCP films as a function of xMMA.

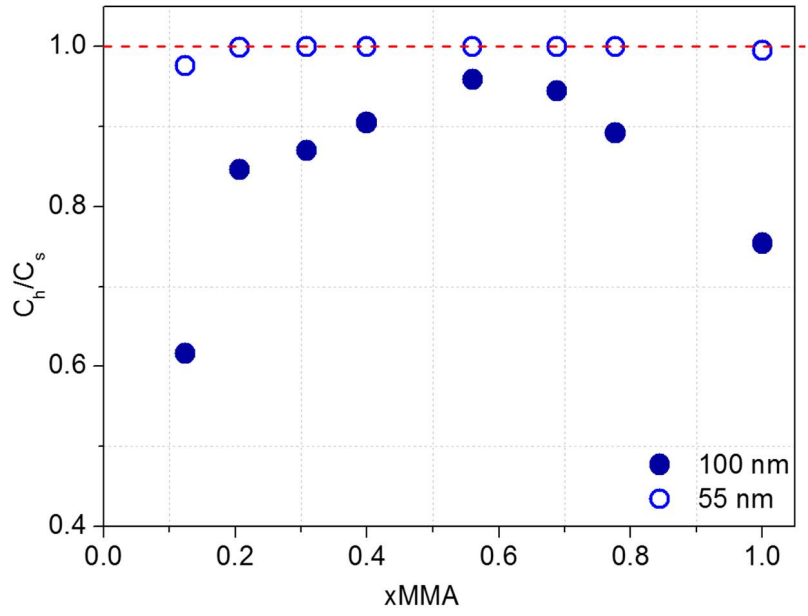

**Figure 8.** Concentration ratio as function of xMMA calculated with the diffusion coefficients reported in Fig. 7 for 60 sec of exposure to TMA vapor.

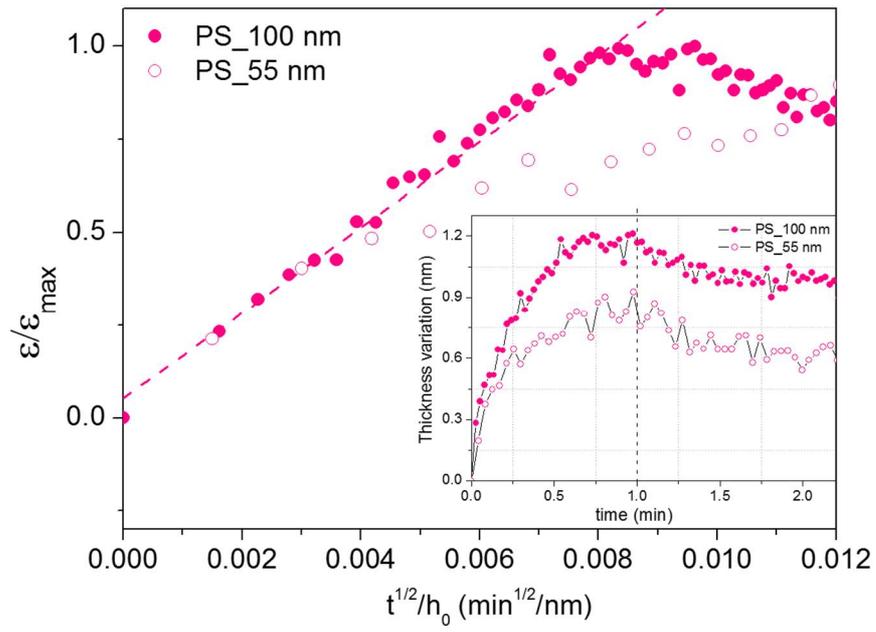

**Figure 9.** $\varepsilon(t)/\varepsilon_{max}$ during the 1st TMA exposure phase as a function of $\sqrt{t}/h_0$ for PS samples. In the inset: thickness variation of PS films during first cycle of TMA exposure and purge.

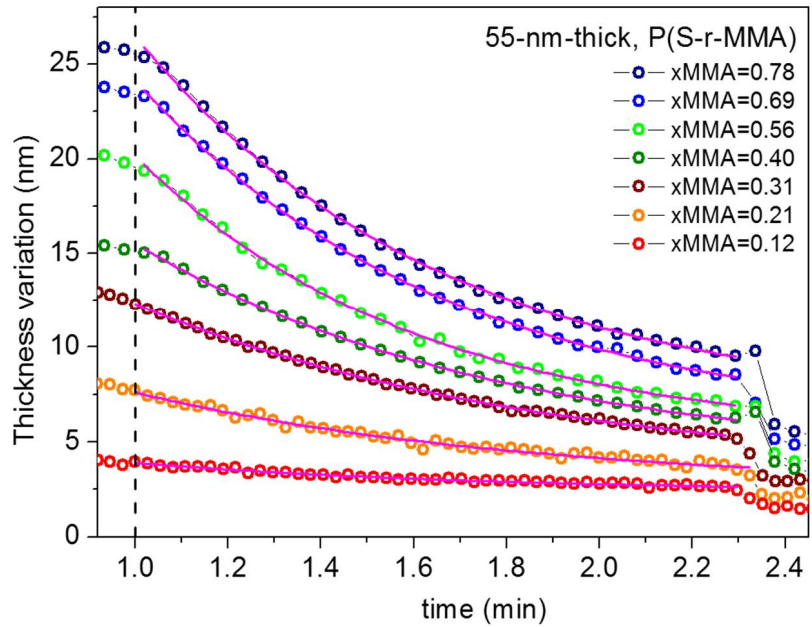

**Figure 10**. *In-situ* SE measured thickness of 55 nm thick P(S-r-MMA) samples during the 1st TMA purging phase. The dashed line divides the TMA exposure phase and the TMA purging phase. An exponential model is used to fit the data to extract the de-swelling time constants ($\tau$).

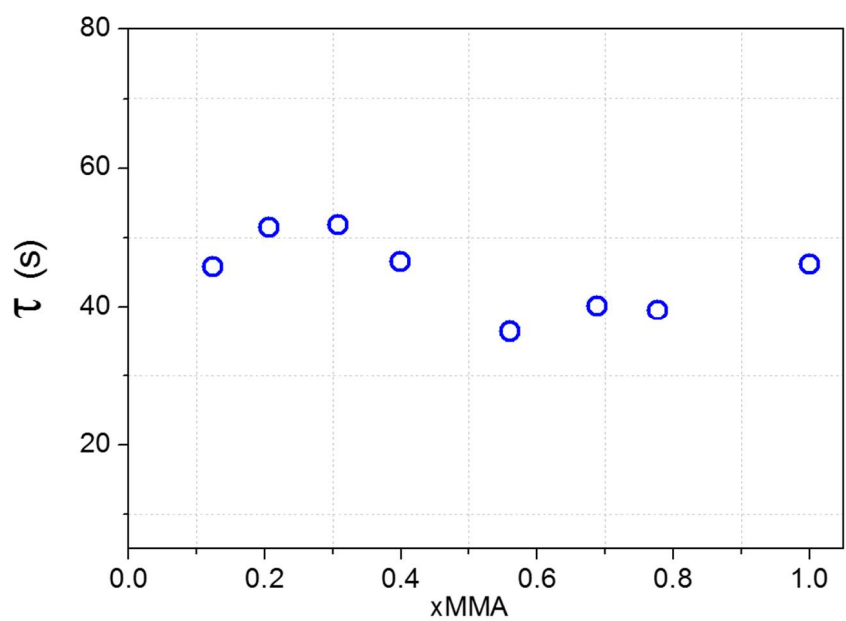

**Figure 11.** Desorption time constant (τ) as a function of MMA fraction present in infiltrated films calculated for samples of 55 nm. It seems to be independent from the MMA fraction within the polymeric film.

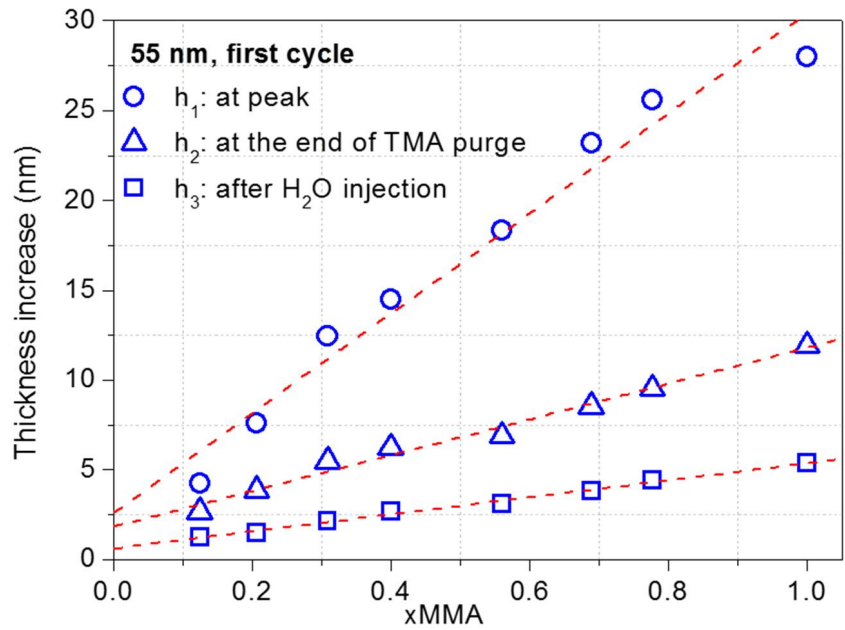

**Figure 12.** Thickness increase at peak ($h_1$), before ($h_2$) and after ($h_3$) $H_2O$ injection, during the first SIS cycle of 55 nm thick samples as a function of MMA fraction. The 55 nm thick case presents a linear growth with the MMA fraction within the polymer films.

**Table 1.** Molar Mass (Mn), Polydispersity Index (Đ), volume fraction of MMA units (%) and fraction of MMA units (xMMA) of PS, PMMA and P(S-r-MMA) RCPs.

| Polymer | % MMA | $M_n$ (g·mol$^{-1}$) | Đ | xMMA |
|---|---|---|---|---|
| **PS** | 0 | 13000 | 1.10 | 0 |
| **P(S-r-MMA)** | 12 | 11600 | 1.19 | 0.12 |
| **P(S-r-MMA)** | 20 | 10600 | 1.18 | 0.21 |
| **P(S-r-MMA)** | 30 | 12300 | 1.32 | 0.31 |
| **P(S-r-MMA)** | 39 | 14200 | 1.25 | 0.40 |
| **P(S-r-MMA)** | 55 | 14800 | 1.52 | 0.56 |
| **P(S-r-MMA)** | 68 | 12500 | 1.27 | 0.69 |
| **P(S-r-MMA)** | 77 | 13100 | 1.36 | 0.78 |
| **PMMA** | 100 | 14000 | 1.21 | 1 |

**ToC**

# Effect of the Density of Reactive Sites in P(S-r-MMA) Film During $Al_2O_3$ Growth by Sequential Infiltration Synthesis

Federica E. Caligiore, Daniele Nazzari, Elena Cianci*, Katia Sparnacci, Michele Laus, Michele Perego*, Gabriele Seguini.

The amount of alumina grown in the P(S-$r$-MMA) films by sequential infiltration synthesis was analyzed by spectroscopic ellipsometry and it depends on MMA content within random copolymer films. A relatively low concentration of MMA in the copolymer matrix is enough to guarantee the volumetric growth of alumina in the polymer film.

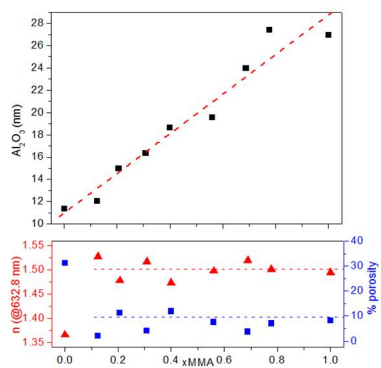